\documentstyle[preprint,aps,epsf,floats]{revtex}

\tighten

\def\OMIT#1{{}}

\def\Dslash{D\hskip-0.65em /}

\def\cb{{\cal B}}
\def\cbb{{\overline{\cal B}}}
\def\cf{{\cal F}}

\begin{document}

\preprint{\vbox{
\hbox{NT@UW-03-014}
\hbox{INT-PUB-03-010}
}}

\title{Nucleon Properties at Finite Lattice Spacing\\
in Chiral Perturbation Theory}
\author{{\bf Silas R.~Beane}$^a$  and {\bf Martin J.~Savage}$^{b}$}
\address{$^a$ Institute for Nuclear Theory, University of Washington, \\
Seattle, WA 98195. }
\address{$^b$ Department of Physics, University of Washington, \\
Seattle, WA 98195. }

\maketitle

\begin{abstract} 
Properties of the proton and neutron are studied in partially-quenched
chiral perturbation theory at finite lattice spacing. Masses, magnetic
moments, the matrix elements of isovector twist-2 operators
and axial-vector currents are examined at the one-loop level in a double
expansion in the light-quark masses and the lattice spacing. This work
will be useful in extrapolating the results of simulations 
using Wilson valence and
sea quarks, as well as simulations using Wilson sea quarks and Ginsparg-Wilson
valence quarks, to the continuum.
\end{abstract}

\bigskip
\vskip 8.0cm
\leftline{June 2003}

\vfill\eject

\section{Introduction}

Impressive progress is currently being made in understanding
properties and interactions of the low-lying hadrons using lattice
QCD.  However, computational limitations necessitate the use of quark
masses, $m_q$, that are significantly larger than those of nature, lattice
spacings, $a$, that are not significantly smaller than the physical scale of
interest, and lattice sizes, $L$, that are not significantly larger than
the physical scale of interest. In order to make a connection between
lattice QCD calculations of the foreseeable future and nature,
extrapolations in the quark masses, lattice spacing and lattice
volume are required. 
Assuming a  hierarchy of mass scales,
\begin{equation}
L^{-1}\ll m_q \ll \Lambda_\chi \ll a^{-1}\ \ ,
\label{eq:hierarchy}
\end{equation}
where $\Lambda_\chi$ is the scale of chiral symmetry breaking (a typical
QCD scale), and working in the infinite volume limit,
the appropriate tool for incorporating the light quark masses and the finite
lattice spacing into hadronic observables is effective field theory (EFT).

Chiral perturbation theory ($\chi$PT) provides
a systematic description of low-energy QCD near the chiral limit
and is therefore an EFT which exploits the hierarchy  $m_q \ll \Lambda_\chi$.
This technology has been extended to 
describe both quenched QCD
(QQCD)~\cite{Sharpe90,S92,BG92,LS96,S01a} with quenched chiral
perturbation theory (Q$\chi$PT) and partially-quenched QCD
(PQQCD)~\cite{Pqqcd1,Pqqcd2,Pqqcd3,Pqqcd4,SS01} with partially-quenched chiral perturbation
theory (PQ$\chi$PT).  It is hoped that future lattice simulations can
be performed with sufficiently small quark masses where the chiral
expansion is convergent, and can be used to extrapolate down to the
quark masses of nature. Recently, meson and baryon properties have
been studied extensively in both Q$\chi$PT~\cite{LS96,S01a} and
PQ$\chi$PT~\cite{CSn,BSn,BSpv,Leinweber:2002qb}. The effective field
theory (EFT) describing the low-energy dynamics of two-nucleon systems
in PQQCD has also been explored~\cite{BSnn,ABSl}.

In order to construct the low-energy EFT at finite lattice spacing,
written in terms of the hadronic fields, one must first construct the
underlying lattice theory, written in terms of the quark and gluon
fields.  The lattice theory and the continuum theory coincide in the
$a=0$ limit, but away from this limit the
theories differ. The lattice breaks the Lorentz group ($O(4)$ in
Euclidean space) down to the discrete symmetry group of the lattice,
which we will take to be the symmetry group $H(4)$ of a
hypercubic lattice. As first discussed by Symanzik~\cite{Symanzik}, 
the strong-interaction Lagrange density at $a\ne 0$
will receive contributions from an infinite series of operators,
$\sim\sum a^k {\cal O}^{(4+k)}$.  Therefore, the contribution from
terms of ${\cal O}(a^n)$ to a given strong-interaction observable will,
according to eq.~(\ref{eq:hierarchy}),
be suppressed by factors of $\sim a^n\Lambda_\chi^n$. For Wilson
fermions~\cite{wilson}, where chiral symmetry is not a good symmetry, it is
straightforward to show that at ${\cal O}(a)$ the Symanzik Lagrange
density has the form, 
once appropriate redefinitions and renormalizations have been performed,
\begin{eqnarray}
{\cal L} & = & \overline{\psi}\left(\Dslash + m_q\right) \psi
\ +\ a c_{sw} \overline{\psi}\sigma^{\mu\nu}G_{\mu\nu}\psi
\ +\ ...
\end{eqnarray}
where $c_{sw}$ is the Sheikholeslami-Wohlert~\cite{SWterm} coefficient
that must be determined numerically. 
For lattice fermions that satisfy the Ginsparg-Wilson (GW)
condition~\cite{Ginsparg:1981bj}, 
such as Kaplan fermions~\cite{KaplanF} and overlap fermions~\cite{Narayanan:ss}, 
where chiral symmetry is a good symmetry, 
the coefficient of the Sheikholeslami-Wohlert~\cite{SWterm} term vanishes,
$c_{sw}=0$. The power-counting
we will use in this work treats both $m_q$ and the lattice spacing
$a$ as small.  The small dimensionless parameters that we will use in
our low-energy EFT are
\begin{eqnarray}
p^2 & \sim & {m_q\over\Lambda_\chi}\ \ \sim\ \ a\Lambda_\chi
\ \ \sim\ \ {\partial^2\over\Lambda_\chi^2}
\ \ \ ,
\end{eqnarray}
where $\partial$ represents the derivative operator.

Following earlier work of Sharpe and Singleton~\cite{Sharpe:1998xm} (see also
Ref.~\cite{Lee:1999zx}), Rupak and Shoresh~\cite{RSa} have 
extended $\chi$PT to ${\cal O}(p^4)$ including the effects of a finite lattice spacing
at ${\cal O}(a)$ for Wilson fermions, and computed the Goldstone-boson masses.
Together with
B\"ar~\cite{BRSa} these same authors have generalized the results to ``mixed'' actions where
different types of lattice fermions are used for the sea and valence
quarks.  Recently, this work was extended 
to ${\cal O}(a^2)$ for both Wilson and mixed actions~\cite{BRSaa,aoki}.
The pion decay constant has also been computed
to ${\cal O}(a^2)$ for Wilson fermions in QCD by Aoki~\cite{aoki}.
When considering the matrix elements of operators coupled to external
sources, such as the axial-vector current matrix elements or the matrix
elements of twist-2 operators, there are contributions at ${\cal
O}(a)$ from the operator itself, in addition to the
Sheikholeslami-Wohlert term in the strong-interaction sector. In this work we
consider the leading ${\cal O}(a)$ corrections to nucleon properties.
We compute the contributions to the nucleon masses at
${\cal O}(p^{3})$, to their magnetic moments at ${\cal O}(p)$, to
their isovector axial-vector matrix elements at ${\cal O}(p^2)$ and to the
matrix element of the $n=2$ isovector twist-2 operator at ${\cal
O}(p^2)$.

%%%%%%%%%%%%%%%%%  PQQCD %%%%%%%%%%%%%%%
\section{PQQCD at Finite Lattice Spacing}

The Symanzik effective Lagrange density at ${\cal O}(p)$ which describes
the quark-sector of PQQCD for two light flavors is
\begin{eqnarray}
{\cal L} & = & 
\overline{Q}\ 
\left[\ i\Dslash -m_{Q}\ \right]\ Q
\ +\ 
\overline{Q}\ \sigma^{\mu\nu}G_{\mu\nu} \ A_Q\  Q
\ \ \ ,
\label{eq:PQQCD}
\end{eqnarray}
where the valence, sea, and ghost quarks are combined into the
column vector
\begin{eqnarray}
Q & = & \left(u,d,j,l,\tilde u,\tilde d\right)^T
\ \ \ .
\label{eq:quarkvec}
\end{eqnarray}
The $u$ and $d$ are valence quarks, the $\tilde u$ and $\tilde d$
are ghost quarks, and the $j$ and $l$ are sea quarks.  The
mass matrix, $m_Q$, is $m_Q={\rm diag}(m_u , m_d , m_j , m_l , m_u ,
m_d)$, and the Sheikholeslami-Wohlert coefficient matrix is $A_Q = a \
{\rm diag}(c_{sw}^{(V)} , c_{sw}^{(V)} , c_{sw}^{(S)} , c_{sw}^{(S)} ,
c_{sw}^{(V)} , c_{sw}^{(V)})$~\cite{RSa,BRSa}. As mentioned previously, when both the
valence and sea quarks are Wilson fermions $c_{sw}^{(V)} =
c_{sw}^{(S)}$, but when the valence quarks are GW fermions while
the sea quarks are Wilson fermions, $c_{sw}^{(V)}=0$.

The graded equal-time commutation relations for two fields are
\begin{eqnarray}
Q_i^\alpha ({\bf x}) Q_k^{\beta \dagger} ({\bf y}) - 
(-)^{\eta_i\eta_k}Q_k^{\beta \dagger} ({\bf y})Q_i^\alpha ({\bf x})
& = & 
\delta^{\alpha\beta}\delta_{ik}\delta^3({\bf x}-{\bf y})
\ \ \ ,
\label{eq:comm}
\end{eqnarray}
where $\alpha,\beta$ are spin-indices and $i,k$ are flavor indices.
The objects $\eta_k$ correspond to the parity of the component of
$Q_k$, with $\eta_k=+1$ for $k=1,2,3,4$ and $\eta_k=0$ for $k=5,6$,
and the graded commutation relations for two $Q$'s or two
$Q^\dagger$'s are analogous.  The left- and right-handed quark fields,
$Q_{L,R}$ in eq.~(\ref{eq:quarkvec}), transform in the fundamental
representation of $SU(4|2)_{L,R}$, respectively.  The ground floor of
$Q_L$ transforms as a $({\bf 4},{\bf 1})$ of $SU(4)_{qL}\otimes
SU(2)_{\tilde q L}$ while the first floor transforms as $({\bf 1},{\bf
2})$, and the right handed field $Q_R$ transforms analogously.  In the
absence of the quark-mass and Sheikholeslami-Wohlert terms,
$m_Q=A_Q=0$, the Lagrange density in eq.~(\ref{eq:PQQCD}) has a graded
symmetry $U(4|2)_L\otimes U(4|2)_R$, where the left- and right-handed
quark fields transform as $Q_L\rightarrow U_L Q_L$ and $Q_R\rightarrow
U_R Q_R$, respectively.  The strong anomaly reduces the symmetry of the
theory to $SU(4|2)_L\otimes SU(4|2)_R\otimes
U(1)_V$~\cite{SS01}.  It is assumed that this symmetry is
spontaneously broken according to the pattern $SU(4|2)_L\otimes SU(4|2)_R\otimes
U(1)_V\rightarrow SU(4|2)_V\otimes U(1)_V$ so that an identification
with QCD can be made.

%%%%%%%%%%%%  Mesons  %%%%%%%%%%%%%%%%%
\subsection{The Pseudo-Goldstone Bosons}

In order to construct the Lagrange density describing the dynamics of
the pseudo-Goldstone bosons at ${\cal O}(p^2)$, we allow $m_Q$ and
$A_Q$ to transform under the graded chiral group~\cite{Sharpe:1998xm,RSa,BRSa}.
This leads to
\begin{eqnarray}
{\cal L } & = & 
{f^2\over 8} 
{\rm str}\left[\ \partial^\mu\Sigma^\dagger\partial_\mu\Sigma\ \right]
 \ +\ 
\lambda_M\ {f^2\over 4} 
{\rm str}\left[\ m_Q\Sigma^\dagger + m_Q\Sigma\ \right]
\ +\ 
\lambda_A\ {f^2\over 4} 
{\rm str}\left[\ A_Q\Sigma^\dagger + A_Q\Sigma\ \right]
\nonumber\\
&& \ +\ \alpha_\Phi\partial^\mu\,\Phi_0 \partial_\mu\,\Phi_0\ -\ m_0^2\Phi_0^2
\ \ \ \ ,
\label{eq:lagpi}
\end{eqnarray}
where $\alpha_\Phi$ and $m_0$ are quantities that do not vanish in the chiral limit.
The meson field is incorporated in $\Sigma$ via
\begin{eqnarray}
\Sigma & = & \exp\left({2\ i\ \Phi\over f}\right)
\ =\ \xi^2
\ \ \ ,\ \ \ 
\Phi \ =\  \left(\matrix{ M &\chi^\dagger \cr \chi &\tilde{M} }\right)
\ \ \ ,
\label{eq:phidef}
\end{eqnarray}
where $M$ and $\tilde M$ are matrices containing bosonic mesons while
$\chi$ and $\chi^\dagger$ are matrices containing fermionic mesons,
with
\begin{eqnarray}
M & = & 
\left(\matrix{
\eta_u & \pi^+ & J^0 & L^+ \cr
\pi^- & \eta_d & J^- & L^0\cr
\overline{J}^0 & J^+ & \eta_j & Y_{jl}^+\cr
L^- & \overline{L}^0 & Y_{jl}^- & \eta_l  }
\right)
\ ,\ 
\tilde M \ =\  \left(\matrix{\tilde\eta_u & \tilde\pi^+ \cr
\tilde\pi^- & \tilde\eta_d
}\right)
\ ,\
\chi \ =\ 
\left(\matrix{\chi_{\eta_u} & \chi_{\pi^+} &  
\chi_{J^0} & \chi_{L^+}\cr
\chi_{\pi^-} & \chi_{\eta_d} & 
\chi_{J^-} & \chi_{L^0} }
\right)
\ .
\label{eq:mesdef}
\end{eqnarray}
The convention we use
corresponds to $f~\sim~132~{\rm MeV}$, and the charge assignments have
been made using an electromagnetic charge matrix of ${\cal Q}^{(PQ)} =
{1\over 3} {\rm diag}\left(2,-1,2,-1,2,-1\right)$.  The singlet field
is defined to be $\Phi_0 ={\rm str}\left(\ \Phi\ \right)/\sqrt{2}$,
and its mass $m_0$ can be taken to be of order the scale of chiral
symmetry breaking, $m_0\rightarrow\Lambda_\chi$~\cite{SS01}. Hence the
parameters $\alpha_\Phi$ and $m_0$ decouple from the low-energy EFT in 
PQ$\chi$PT~\cite{SS01}. 

It is straightforward to show that the meson masses resulting from
this Lagrange density are
\begin{eqnarray}
m_{ud}^2 & = & \lambda_M \ (m_u+m_d)\ +\ 2 \lambda_A\  a\  c_{sw}^{(V)} \ \ ,
\nonumber\\
m_{uu}^2 & = &  
2\  \lambda_M \ m_u\ +\ 2\  \lambda_A\  a\  c_{sw}^{(V)} \ \ ,
\nonumber\\
m_{ju}^2 & = & 
\lambda_M \ (m_j+m_u)\ +\  \lambda_A ( a\  c_{sw}^{(V)}\  +\  a\  c_{sw}^{(S)}) \ \ ,
\nonumber\\
m_{jl}^2 & = & 
\lambda_M \ (m_j+m_l)\ +\  2\lambda_A  a\  c_{sw}^{(S)}\ \ ,
\label{eq:mesmasses}
\end{eqnarray}
and so forth, where $m^2_{ab}$ denotes the mass-squared of a meson
containing a (anti-) quark of flavor $a$ and one of flavor $b$ (either
valence, sea or ghost).  The meson masses have been computed out to
${\cal O}(m_q^2 a)$ in Refs.~\cite{RSa,BRSa}.

%%%%%%%%%%%%% N Delta  %%%%%%%%%%%%%%%%%%%%%
\subsection{The Nucleons and $\Delta$-Resonances}

The free Lagrange density for the ${\bf 70}$-dimensional baryon
supermultiplet, ${\cal B}_{ijk}$, containing the nucleon and for the
${\bf 44}$-dimensional baryon supermultiplet, ${\cal T}^\mu_{ijk}$,
containing the $\Delta$-resonances is~\cite{CSn,BSn}, at leading order
(${\cal O}(p)$),
\begin{eqnarray}
{\cal L} & = & 
i\left(\overline{\cal B} v\cdot {\cal D} {\cal B}\right)
+2\alpha_M^{\rm (PQ)} \left(\overline{\cal B}{\cal B}{\cal M}_+\right)
+2\beta_M^{\rm (PQ)} \left(\overline{\cal B}{\cal M}_+{\cal B}\right)
+2\sigma_M^{\rm (PQ)} \left(\overline{\cal B}{\cal B}\right)\ 
{\rm str}\left({\cal M}_+\right)
\nonumber\\
& + & 
2\alpha_A^{\rm (PQ)} \left(\overline{\cal B}{\cal B}{\cal A}_+\right)
+2\beta_A^{\rm (PQ)} \left(\overline{\cal B}{\cal A}_+{\cal B}\right)
+2\sigma_A^{\rm (PQ)} \left(\overline{\cal B}{\cal B}\right)\ 
{\rm str}\left({\cal A}_+\right)
\nonumber\\
& - & 
i \left(\overline{\cal T}^\mu v\cdot {\cal D} {\cal T}_\mu\right)
+ 
\Delta\ \left(\overline{\cal T}^\mu {\cal T}_\mu\right)
-2\gamma_M^{\rm (PQ)}
 \left(\overline{\cal T}^\mu{\cal M}_+{\cal T}_\mu\right)
-2 \overline{\sigma}_M^{\rm (PQ)}
  \left(\overline{\cal T}^\mu {\cal T}_\mu\right)\
{\rm str}\left({\cal M}_+\right)
\nonumber\\
& -& 
2\gamma_A^{\rm (PQ)}
 \left(\overline{\cal T}^\mu{\cal A}_+{\cal T}_\mu\right)
-2 \overline{\sigma}_A^{\rm (PQ)}
  \left(\overline{\cal T}^\mu {\cal T}_\mu\right)\
{\rm str}\left({\cal A}_+\right)
\ ,
\label{eq:free}
\end{eqnarray}
where $\Delta$ is the mass splitting between the ${\bf 70}$ and ${\bf
44}$, ${\cal M}_+={1\over 2}\left(\xi^\dagger m_Q\xi^\dagger + \xi
m_Q\xi\right)$, $\xi=\sqrt{\Sigma}$, and ${\cal A}_+={1\over
2}\left(\xi^\dagger A_Q\xi^\dagger + \xi A_Q\xi\right)$.  The terms
that arise at ${\cal O}(p)$ from the
Sheikholeslami-Wohlert~\cite{SWterm} operator have coefficients
$\alpha_A^{\rm (PQ)}$, $\beta_A^{\rm (PQ)}$, $\sigma_A^{\rm (PQ)}$,
$\gamma_A^{\rm (PQ)}$, and $\overline{\sigma}_A^{\rm (PQ)}$.

The Lagrange density describing the interactions of the ${\bf 70}$ and
${\bf 44}$ with the pseudo-Goldstone bosons at LO in the chiral
expansion is~\cite{LS96},
\begin{eqnarray}
{\cal L} & = & 
2\alpha\ \left(\overline{\cal B} S^\mu {\cal B} A_\mu\right)
\ +\ 
2\beta\ \left(\overline{\cal B} S^\mu A_\mu {\cal B} \right)
\ +\  
2{\cal H} \left(\overline{\cal T}^\nu S^\mu A_\mu {\cal T}_\nu \right)
\nonumber\\
& &  
\ +\ 
\sqrt{3\over 2}{\cal C} 
\left[\ 
\left( \overline{\cal T}^\nu A_\nu {\cal B}\right)\ +\ 
\left(\overline{\cal B} A_\nu {\cal T}^\nu\right)\ \right]
\ ,
\label{eq:ints}
\end{eqnarray}
where $S^\mu$ is the covariant
spin-vector~\cite{JMheavy,JMaxial,Jmass}.  Restricting ourselves to
the valence sector, we can compare eq.~(\ref{eq:ints}) with the
LO interaction Lagrange density of QCD,
\begin{eqnarray}
{\cal L} & = & 
2 g_A\  \overline{N} S^\mu  A_\mu N
\ +\ 
g_1\overline{N} S^\mu N\ {\rm tr}
\left[\ A_\mu\ \right]
\ +\  
g_{\Delta N}\ 
\left[\ 
\overline{T}^{abc,\nu}\  A^d_{a,\nu}\  N_b \ \epsilon_{cd} 
\ +\ {\rm h.c.}
\ \right]
\nonumber\\
& & 
+\ 
2 g_{\Delta\Delta}\  
\overline{T}^\nu S^\mu A_\mu T_\nu 
\ +\ 
2 g_{X}\  
\overline{T}^\nu S^\mu \ T_\nu \  {\rm tr}
\left[\ A_\mu\ \right]
\ ,
\label{eq:intsQCD}
\end{eqnarray}
and find that at tree-level
\begin{eqnarray}
\alpha & = & {4\over 3} g_A\ +\ {1\over 3} g_1
\ \ \ ,\ \ \ 
\beta \ =\ {2\over 3} g_1 - {1\over 3} g_A
\ \ \ ,\ \ \ 
{\cal H} \ =\ g_{\Delta\Delta}
\ \ \ ,\ \ \ 
{\cal C} \ =\ -g_{\Delta N}
\ \ \ ,
\label{eq:axrels}
\end{eqnarray}
with $g_X=0$. Considering only the nucleons, and decomposing the
Lagrange density in eq.~(\ref{eq:intsQCD}) into the mass eigenstates
of the isospin-symmetric limit, $\pi^{\pm}, \pi^0$ and $\eta$, we have
\begin{eqnarray}
{\cal L} & = & 
2 g_A\  \overline{N} S^\mu  \tilde A_\mu N
\ +\ 
{\sqrt{2}\over f}\ 
\left(g_A+g_1\right) \overline{N} S^\mu N\ \partial_\mu\eta
\ \ ,
\end{eqnarray}
where $\tilde A_\mu$ is the axial-vector field of pions only
(excluding the isosinglet meson).  In the isospin-symmetric limit,
with the mass of the $\eta$ being of order $\sim\Lambda_{\chi}$, all
expressions must be independent of the coupling $g_1$.  At the order
we work to in this paper, higher-order interactions do not contribute.

%%%%%%%%%%% Masess  %%%%%%%%%%%%%%
\section{Nucleon Masses}

The mass of the $i$-th baryon in the ${\bf 70}$-dimensional 
baryon supermultiplet has an expansion in $m_q$ and $a$, of the form
\begin{eqnarray}
M_i & = & M_0(\mu)\ -\ M_i^{(1)}(\mu)\ -\ M_i^{(3/2)}(\mu)\ +\ ...
\ \ \ ,
\label{eq:massexp}
\end{eqnarray}
and we will be interested only in the proton and neutron masses,
i.e. $i=p,n$.  The superscript denotes the order in the expansion,
i.e. $M_p^{(3/2)}(\mu)$ denotes a contribution of ${\cal O}(p^3)$.
The term $M_0(\mu)$ is the same for all baryons in the
supermultiplet, and is non-zero in the $m_Q , a\rightarrow 0$ limits.
The $a=0$ values of $M_i^{(1)}(\mu)$ and $M_i^{(3/2)}(\mu)$ can be
found in Ref.~\cite{BSn} for arbitrary quark masses.  For $a\ne 0$ we
find that $M_i^{(1)}(\mu)$ becomes
\begin{eqnarray}
M_p^{(1)} & = & 
{1\over 3} m_u \left(5\alpha_M^{(PQ)} + 2\beta_M^{(PQ)}\right)
+{1\over 3}  m_d \left(\alpha_M^{(PQ)} + 4\beta_M^{(PQ)}\right)
+2 \sigma_M^{(PQ)}\left(m_j+m_l\right) 
\nonumber\\
& & 
\ +\  2 \ a\  c_{sw}^{(V)}\  \left(\alpha_A^{(PQ)} + \beta_A^{(PQ)}\right)
\ +\  4 \ \sigma_A^{(PQ)}\  a\  c_{sw}^{(S)} \ \ ,
\nonumber\\
M_n^{(1)} & = & 
{1\over 3} m_u \left(\alpha_M^{(PQ)} + 4\beta_M^{(PQ)}\right)
+{1\over 3}  m_d \left(5\alpha_M^{(PQ)} + 2\beta_M^{(PQ)}\right)
+2 \sigma_M^{(PQ)}\left(m_j+m_l\right)
\nonumber\\
& &
\ +\  2\  a\  c_{sw}^{(V)}\  \left(\alpha_A^{(PQ)} + \beta_A^{(PQ)}\right)
\ +\  4 \ \sigma_A^{(PQ)}\  a\  c_{sw}^{(S)}
\ \ \ .
\end{eqnarray}
At this order, 
the finite $a$ contributions are the same for the proton and neutron.
This has to be the case as the lattice spacing, being the same for the
$u$ and $d$-quarks, transforms as an isoscalar.  The
$M_i^{(3/2)}(\mu)$ contributions, arising from one-loop diagrams in
PQ$\chi$PT, have exactly the same form as in eqs.(38) and (40) of
Ref.~\cite{BSn}, however, it is understood that the meson masses are
evaluated with the relations in eq.~(\ref{eq:mesmasses}), and hence
have implicit dependence on the lattice spacing.  We see that this
introduces non-analytic dependence on the lattice spacing in the chiral limit.

As the explicit expressions for the loop contributions are quite long, we
will simply quote the results
in the isospin limit (i.e. the sea quarks are degenerate, but different in mass
from the degenerate valence quarks and ghosts),
\begin{eqnarray}
M_p^{(3/2)} & = &  
{1\over 8\pi f^2}\left(\ 
{g_A^2\over 12}\left[ 9 m_{SS}^2 m_{VV} + 16 m_{SV}^3 - 7 m_{VV}^3 \right]
\ + \
{g_1^2\over 12}\left[9 m_{SS}^2 m_{VV} + 10 m_{SV}^3 - 19 m_{VV}^3 \right]
\right. \nonumber\\ && \left.\qquad
\ +\ 
{g_A g_1\over 6}\left[9 m_{SS}^2 m_{VV} + 4 m_{SV}^3 - 13 m_{VV}^3 \right]
\ +\ 
{ 2 g_{\Delta N}^2\over 3\pi} \left[ F_{VV} + F_{SV} \right]
\right)
\ \ \ ,
\end{eqnarray}
where $S$ denotes a sea quark and $V$ denotes a valence quark.
The function $F_{c}=F( m_{c},\Delta,\mu)$ is
\begin{eqnarray}
F (m,\Delta,\mu) & = & 
\left(m^2-\Delta^2\right)\left(
\sqrt{\Delta^2-m^2} \log\left({\Delta -\sqrt{\Delta^2-m^2+i\epsilon}\over
\Delta +\sqrt{\Delta^2-m^2+i\epsilon}}\right)
-\Delta \log\left({m^2\over\mu^2}\right)\ \right)
\nonumber\\
& - & {1\over 2}\Delta m^2 \log\left({m^2\over\mu^2}\right)
\ \ \ ,
\label{eq:massfun}
\end{eqnarray}
where $\mu$ is the renormalization scale, and the meson masses 
depend upon light-quark masses and the lattice spacing through
eq.~(\ref{eq:mesmasses}). Note that we use dimensional regularization with 
$\overline{MS}$ to regulate divergent integrals.
Of course, all regulators must give equivalent results.

In the QCD  and  isospin limit, $m_j, m_l, m_u,
m_d \rightarrow \overline{m}$, while at 
finite $a$, the nucleon mass is
\begin{eqnarray}
M_N & = & M_0 
- 2 \ \overline{m}\ \left( \alpha_M^{(PQ)} + \beta_M^{(PQ)} + 2\sigma_M^{(PQ)}\right)
- {1\over 8\pi f^2}\left[\ {3\over 2} g_A^2 m_\pi^3
\ +\ {4 g_{\Delta N}^2\over 3\pi} F_\pi\ \right]
\nonumber\\
&& 
\ -\ 2 \ a\  c_{sw}^{(V)}\ 
\left( \alpha_A^{(PQ)} + \beta_A^{(PQ)} \right)
\ -\ 4\ a\  c_{sw}^{(S)}\ \sigma_A^{(PQ)}
\ \ \ .
\end{eqnarray}
%

%%%%%%%%%%%%%%%  Mags  %%%%%%%%%%%%%%%%%%%%
\section{Nucleon Magnetic Moments}

The most general analysis of the nucleon magnetic moments 
requires us to determine the vector-current operator out to the order
in the lattice spacing that we are working.  While the ${\cal O}(a)$
corrections are known~\cite{Capitani:2000xi}, in order to compute the
magnetic moments at ${\cal O}(p^{1/2})$ only the continuum limit of the
vector-current operator is required.  Up to ${\cal O}(p^{1/2})$, it is
convenient to write the magnetic moment of the $i$-th nucleon as
\begin{eqnarray}
\mu_i & = & \alpha_i\ +\ {M_N\over 4\pi f^2}\ \left[\ 
\beta_i\ +\ \beta_i^\prime\ \right]
\ +\ ...
\ \ \ ,
\label{eq:magdef}
\end{eqnarray}
where the constants, $\alpha_i$, $\beta_i$ and $\beta_i^\prime$ in
PQQCD are given in eqs.(49)-(51) in Ref.~\cite{BSn}.  
The only modification required at
finite lattice spacing is to use eq.~(\ref{eq:mesmasses}) in
evaluating the meson masses.  Therefore, we see that the lattice
spacing first appears through the meson masses at order ${\cal
O}(p^{1/2})$,
and hence the leading contribution of the finite lattice spacing
is non-analytic in the lattice spacing in the chiral limit.

In the isospin limit one finds
\begin{eqnarray}
\beta_p & = & 
-{g_A^2\over 9}\left[ 8 m_{SV} + m_{VV}\right]
\ -\ {4 g_A g_1\over 9} \left[ m_{SV} - m_{VV}\right]
\ -\ {g_1^2\over 18} \left[ m_{SV} - m_{VV}\right] 
\nonumber\\
&&\ +\ (q_j+q_l) \left[ m_{SV} - m_{VV}\right] {1\over 3} \left(2g_A^2 + g_A g_1
  + {5\over 4} g_1^2\right) \ \ ,
\nonumber\\
\beta_p^\prime & = & 
{g_{\Delta N}^2\over 9}\left[ -2 {\cal F}_{VV}  
+ {3\over 2} (q_j+q_l) \left({\cal F}_{VV} - {\cal F}_{SV}\right) \right] \ \ ,
\nonumber\\
\beta_n & = & 
{g_A^2\over 9}\left[ 4 m_{SV} + 5 m_{VV}\right]
\ +\ {2 g_A g_1\over 9} \left[ m_{SV} - m_{VV}\right]
\ -\ {2 g_1^2\over 9} \left[ m_{SV} - m_{VV}\right]
\nonumber\\
&&\ +\ (q_j+q_l) \left[ m_{SV} - m_{VV}\right] {1\over 3} \left(2g_A^2 + g_A g_1
  + {5\over 4} g_1^2\right) \ \ ,
\nonumber\\
\beta_n^\prime & = & 
{g_{\Delta N}^2\over 9}\left[ {\cal F}_{VV}  + {\cal F}_{SV}  
+ {3\over 2} (q_j+q_l) \left({\cal F}_{VV} - {\cal F}_{SV}\right) \right]
\ \ \ ,
\label{eq:magmoms}
\end{eqnarray}
where the function $\cf_i={\cal F}(m_i,\Delta,\mu)$ is
\begin{eqnarray}
\pi {\cal F}(m,\Delta,\mu)
& = & \sqrt{\Delta^2-m^2}\log\left({\Delta-\sqrt{\Delta^2-m^2+i\epsilon}
\over \Delta+\sqrt{\Delta^2-m^2+i\epsilon}}\right)
\ -\ \Delta\log\left({m^2\over\mu^2}\right)
\ \ \ .
\label{eq:magfun}
\end{eqnarray}
We have used 
${\cal Q}={\rm diag}\left(+{2\over 3},-{1\over 3},q_j,q_l,q_j,q_l\right)$,
for the electromagnetic charge matrix in PQQCD~\cite{CSn,BSn}.
The charges of the sea quarks and ghosts, $q_j$ and $q_l$, arise due to the fact
that their charge assignments are not unique, only constrained by the
requirement that electromagnetic observables computed in PQQCD reproduce those
of QCD in the QCD limit~\cite{CSn,BSn,GP01a}.
It is clear from the expressions in eq.~(\ref{eq:magmoms}) that this is indeed
the case.

In the isospin-symmetric QCD limit at finite lattice spacing, 
the magnetic moments become~\cite{CP74,JLMS92}
\begin{eqnarray}
\mu_p & = & \mu_0+\mu_1 
- {M_N\over 4\pi f^2}\left[\ g_A^2 \ m_{\pi} 
+ {2\over 9}\ g_{\Delta N}^2\  {\cal F}_{\pi}\ \right] \ \ ,
\nonumber\\
\mu_n & = & \mu_0-\mu_1 
+ {M_N\over 4\pi f^2}\left[\ g_A^2 \ m_{\pi} 
+ {2\over 9}\ g_{\Delta N}^2 \ {\cal F}_{\pi}\ \right]
\ \ \ .
\label{eq:magmomsQCD}
\end{eqnarray}
The isoscalar and isovector magnetic moment contributions from the LO
dimension-5 operators are $\mu_0$ and $\mu_1$, respectively, and are
independent of the lattice spacing.

%%%%%%%%%%%%%%%  Axials  %%%%%%%%%%%%%%%%%%%%
\section{Nucleon Axial Matrix Elements}

The leading effects of a finite lattice spacing on the matrix elements
of the axial-vector current enter at ${\cal O}(p^2)$ from both one-loop
diagrams and from local counterterms.  In addition to the contribution
from the Sheikholeslami-Wohlert term, there is also a contribution
from the ${\cal O}(a)$ corrections to the axial-current operator.  As
discussed in detail in Ref.~\cite{Luscher:1996sc}, there are only two
operator structures that contribute to the axial-current operator at
${\cal O}(a)$, using the notation of Ref.~\cite{Luscher:1996sc},
\begin{eqnarray}
{\cal O}_{7,\mu}^a & = &  \overline{q}\  {\tau}^a \gamma_5
\left(i \stackrel{\leftrightarrow}{D}_{\mu}\right)\  q
\ \ \ \ ,\ \ \ \
{\cal O}_{8,\mu}^a \ = \  \overline{q}\ {\tau}^a \gamma_\mu \gamma_5\ m_q\ 
q
\ \ \ ,
\label{eq:axa}
\end{eqnarray}
in QCD, where $ \stackrel{\leftrightarrow}{D}_{\mu}
=\overrightarrow{D}_\mu -\overleftarrow{D}_\mu$.  The appearance of
$m_q$ in ${\cal O}_{8,\mu}^a$ renders it ${\cal O}(p^4)$ 
in the power-counting, and so we neglect it, leaving the
contribution from ${\cal O}_{7,\mu}^a$.  The extension to PQQCD
is straightforward. 
In analogy with the electromagnetic interaction, the isovector axial-charge
matrix must be extended to PQQCD and the axial charges of the sea quarks and
ghost must be defined as 
the extension is not unique~\cite{GP01a,CSn,BSn}.
Requiring the axial charge matrix to be
supertraceless implies that the most general extension to the leading
operator is, e.g.,
\begin{eqnarray}
\tau^3 & \rightarrow & \overline{\tau}^3\ =\ {\rm diag}\left(1 , -1 , y_j , y_l
  , y_j , y_l \right)
\ \ \ .
\label{eq:axcharge}
\end{eqnarray}
In order to construct the nucleon axial matrix element at leading
order one uses the spurion construction in which
$\overline{\tau}^a_L\rightarrow L \overline{\tau}^a_L L^\dagger$ and
$\overline{\tau}^a_R\rightarrow R \overline{\tau}^a_L R^\dagger$,
where the axial-current operator is decomposed into contributions from
the left- and right-handed quark fields.

At subleading order, ${\cal O}(p^2)$, the contribution to the axial current in
PQQCD is 
\begin{eqnarray}
\delta A_\mu^a & = & 
\overline{Q}\   \overline{\tau}_{A,7}^a\  \gamma_5 \left( i\stackrel{\leftrightarrow}{D}_\mu\right) Q
\ \ \ ,
\end{eqnarray}
where, for example, the matrix $ \overline{\tau}_{A,7}^3$ is
\begin{eqnarray}
\overline{\tau}^3_{A,7} & = & a\ {\rm diag}\left( c_{A7}^{(V)}\  ,\
  -c_{A7}^{(V)}\  ,\  
y_j c_{A7}^{(S)}\  ,\  y_l c_{A7}^{(S)} \ ,\  y_j c_{A7}^{(V)} 
\ ,\  y_l c_{A7}^{(V)}
\ \right)
\ \ \ ,
\end{eqnarray}
for the charge matrix in eq.~(\ref{eq:axcharge}), where $c_{A7}^{(S)}$
and $c_{A7}^{(V)}$ are the coefficients of the axial-current
corrections for the sea quarks and valence quarks, respectively.  Under
chiral transformations the spurion field transforms as
$\overline{\tau}^3_{A,7}\rightarrow L \overline{\tau}^3_{A,7}
R^\dagger$, like the quark mass matrix.

The leading-order contribution to the matrix elements of the axial current
arises from the operators~\cite{CSn,BSn}
\begin{eqnarray}
^{(PQ)}j_{\mu,5}^a
& = &
2\alpha\ \left(\overline{\cal B} S_\mu {\cal B}\ {\overline{\tau}^a_{\xi +}}\right)
\ +\ 
2\beta\ \left(\overline{\cal B} S_\mu\ {\overline{\tau}^a_{\xi +}}{\cal B} \right)
\ +\  
2{\cal H} \left(\overline{\cal T}^\nu S_\mu\ 
{\overline{\tau}^a_{\xi +}}{\cal T}_\nu \right)
\nonumber\\
& &  
\ +\ 
\sqrt{3\over 2}{\cal C} 
\left[\ 
\left( \overline{\cal T}_\mu\ {\overline{\tau}^a_{\xi +}} {\cal B}\right)\ +\ 
\left(\overline{\cal B}\ {\overline{\tau}^a_{\xi +}} {\cal T}_\mu\right)\ \right]
\ \ ,
\label{eq:LOaxialcurrent}
\end{eqnarray}
where 
$\overline{\tau}^a_{\xi+}  =  {1\over 2}\left(\ 
\xi\overline{\tau}^a\xi^\dagger + \xi^\dagger\overline{\tau}^a\xi\ \right)$.

At ${\cal O}(p^2)$ there are three different contributions, that we
denote as $^{(PQ)}\delta j_{\mu,5}^{(1), a}$, $^{(PQ)}\delta
j_{\mu,5}^{(2), a}$ and $^{(PQ)}\delta j_{\mu,5}^{(3), a}$.  The
contribution from a single insertion of the leading ${\cal O}(a)$
corrections to the axial-current operator is
\begin{eqnarray}
^{(PQ)}\delta j_{\mu,5}^{(1), a}
& = & 
2\gamma_{A1}\ \left(\overline{\cal B} S_\mu {\cal B}\ 
{\overline{\tau}^a_{A,7,\xi+}}\right)
\ +\ 
2\gamma_{A2}\ \left(\overline{\cal B} S_\mu\
  {\overline{\tau}^a_{A,7,\xi+}}{\cal B} 
\right)
\ \ \ ,
\end{eqnarray}
where $\overline{\tau}^a_{A,7,\xi+}  =  {1\over 2}\left(\ 
\xi\overline{\tau}^a_{A,7}\xi + \xi^\dagger\overline{\tau}^a_{A,7}\xi^\dagger\
\right)$. The remaining two contributions arise from a single insertion of the light-quark
mass matrix and from the 
Sheikholeslami-Wohlert term, which are of the form
\begin{eqnarray}
^{(PQ)}j_{\mu,5}^{(2,3),a}
& &=
2 \left[\ 
b_{1, \Gamma }\  \cbb^{kji}\ \{\  \overline{\tau}^a_{\xi +}\ ,\ 
\Gamma_+\ \}^n_i\ S_\mu \cb_{njk}
\right.\nonumber\\ & & \left.
+\ 
b_{2, \Gamma }\ (-)^{(\eta_i+\eta_j)(\eta_k+\eta_n)}\ 
\cbb^{kji}\ \{\  \overline{\tau}^a_{\xi +}\ ,\ \Gamma_+\ \}^n_k\ 
 S_\mu \cb_{ijn}
\right.\nonumber\\ & & \left.
+\ 
b_{3, \Gamma }\  (-)^{\eta_l (\eta_j+\eta_n)}\
\cbb^{kji}\  \left(\overline{\tau}^a_{\xi +}\right)^l_i\ 
\left( \Gamma_+\right)^n_j
 S_\mu \cb_{lnk}
\right.\nonumber\\ & & \left.
+\ 
b_{4, \Gamma } \  (-)^{\eta_l \eta_j + 1}\ 
\cbb^{kji}\ \left(  
\left(\overline{\tau}^a_{\xi +}\right)^l_i\ \left( \Gamma_+\right)^n_j
\ +\ \left(\Gamma_+\right)^l_i 
\left(\overline{\tau}^a_{\xi +}\right)^n_j \right)
 S_\mu \cb_{nlk}
\right.\nonumber\\ & & \left.
+\ b_{5, \Gamma }\  (-)^{\eta_i(\eta_l+\eta_j)}\ 
\cbb^{kji} \left(\overline{\tau}^a_{\xi +}\right)^l_j 
\left( \Gamma_+\right)^n_i
 S_\mu \cb_{nlk}
\ +\ b_{6, \Gamma }\  \cbb^{kji}  \left(\overline{\tau}^a_{\xi +}\right)^l_i 
 S_\mu \cb_{ljk}
\ {\rm str}\left( \Gamma_+ \right) 
\right.\nonumber\\ & & \left.
\ +\ b_{7, \Gamma }\  \ (-)^{(\eta_i+\eta_j)(\eta_k+\eta_n)}\ 
\cbb^{kji}  \left(\overline{\tau}^a_{\xi +}\right)^n_k 
 S_\mu \cb_{ijn}
\ {\rm str}\left( \Gamma_+ \right) 
\right.\nonumber\\ & & \left.
\ +\ b_{8, \Gamma }\ \cbb^{kji}\  S_\mu \cb_{ijk} 
\ {\rm str}\left(\overline{\tau}^a_{\xi +}\   \Gamma_+ \right) 
\ \ \right]
\ ,
\label{eq:axcts}
\end{eqnarray}
where $\Gamma={\cal M}$ and ${\cal A}$.

We write the axial matrix elements as~\cite{BSn}
\begin{eqnarray}
\langle N_b |^{(PQ)}j_{\mu,5} | N_a\rangle
& = & 
\left[\ 
\rho_{ab}
\ +\ {1\over 16\pi^2 f^2}
\left(\ 
\eta_{ab}\ -\ \rho_{ab} {1\over 2}\left[\ w_a+w_b\ \right]
\ +\ y_j\ \eta_{ab}^{(j)}
\ +\ y_l\ \eta_{ab}^{(l)}
\ \right)
\right. \nonumber\\
& & \left. 
\ +\ c_{ab}
\ +\ y_j\ c_{ab}^{(j)}
\ +\ y_l\ c_{ab}^{(l)}
\ +\ d_{ab}
\ +\ y_j\ d_{ab}^{(j)}
\ +\ y_l\ d_{ab}^{(l)}
\ \right]
\ 2 \ \overline{U}_b S_\mu U_a
\ ,
\label{eq:axmat}
\end{eqnarray}
where the constants $\rho_{ab}$, $\eta_{ab}$, $\eta_{ab}^{(k)}$, $w_a$, 
$c_{ab}$, $c_{ab}^{(k)}$, can be found in Ref.~\cite{BSn} with 
$b_j\rightarrow b_{j,M}$.
In extending the $\tau^+$ isovector operator from QCD to PQQCD, one can simply
replace the $\tau^3$ in the upper $2\times 2$ block of $\overline{\tau}^3$
with $\tau^+$, as described in Ref.~\cite{BSn}. The meson masses in  $\rho_{ab}$, $\eta_{ab}$, 
$\eta_{ab}^{(k)}$, $w_a$ are understood to be evaluated at finite lattice spacing via 
eq.~(\ref{eq:mesmasses}). The constants $d_{ab}$ and $d_{ab}^{(k)}$ arise from 
$^{(PQ)}\delta j_{\mu,5}^{(1), a}$ and  $^{(PQ)}\delta j_{\mu,5}^{(3), a}$, and are
\begin{eqnarray}
d_{pp} & = & {a c_{sw}^{(V)}\over 3}\left[\ -2 b_{1,A} + 4 b_{2,A} -
  b_{3,A} + b_{4,A} + 2 b_{5,A}\ \right]
\ +\ {2 a c_{sw}^{(S)}\over 3}\left[ 2 b_{7,A}-b_{6,A}\ \right]
\nonumber\\
&& \ +\ {a c_{A7}^{(V)}\over 3}\left(\ 2\gamma_{A,1} - \gamma_{A,2}\ \right) \ \ ,
\nonumber\\
d_{pp}^{(j)} & = & d_{pp}^{(l)}\ =\ d_{nn}^{(j)} \ = \ d_{nn}^{(l)}\ =\ 
b_{8,A} \left( a c_{sw}^{(S)} - a c_{sw}^{(V)}\right) \ \ ,
\nonumber\\
d_{np} & = & -d_{nn}\ =\ d_{pp}
\ \ \ ,\ \ \ 
d_{np}^{(j)} \ = \ d_{np}^{(l)}\ =\ 0
\ \ \ .
\end{eqnarray}

In the isospin-symmetric QCD limit at finite lattice spacing,
the proton isovector axial matrix element is
\begin{eqnarray}
\langle p |j_{\mu,5}^{\, 3} | p\rangle
 & = & g_A\ - \ {1\over 8\pi^2 f^2}\left(\ 
g_A\left(1+2g_A^2\right)\ L_\pi 
+ (2g_A + {50\over 81}g_{\Delta\Delta})g_{\Delta N}^2 \ J_\pi 
-{16\over 9} g_A g_{\Delta N}^2\ K_\pi 
\right)
\nonumber\\
&& +\ {\overline{m}\over 3}\left(
-2 b_{1,M} + 4 b_{2,M} - b_{3,M} + b_{4,M} + 2 b_{5,M} 
- 2 b_{6,M} + 4 b_{7,M} \right)
\nonumber\\
&& + {a c_{sw}^{(V)}\over 3}\left[\ -2 b_{1,A} + 4 b_{2,A} -
  b_{3,A} + b_{4,A} + 2 b_{5,A}\ \right]
\ +\ {2 a c_{sw}^{(S)}\over 3}\left[ 2 b_{7,A}-b_{6,A}\ \right]
\nonumber\\
&& \ +\ {a c_{A7}^{(V)}\over 3}\left(\ 2\gamma_{A,1} - \gamma_{A,2}\ \right)
\ +\ 
(y_j+y_l) b_{8,A} \left( a c_{sw}^{(S)} -  a c_{sw}^{(V)}\right)
\ \ \ ,
\label{eq:qcdaxials}
\end{eqnarray}
where $L_\pi = m_\pi^2\log\left({m_\pi^2/\mu^2}\right)$, 
$J_\pi=~J(m_\pi, \Delta, \mu)$ is
\begin{eqnarray}
J(m,\Delta,\mu) & = & 
\left(m^2-2\Delta^2\right)\log\left({m^2\over\mu^2}\right)
+2\Delta\sqrt{\Delta^2-m^2}
\log\left({\Delta-\sqrt{\Delta^2-m^2+ i \epsilon}\over
\Delta+\sqrt{\Delta^2-m^2+ i \epsilon}}\right)
\ \ \ ,
\label{eq:decfun}
\end{eqnarray}
and $K_\pi~=~K(m_\pi,\Delta,\mu)$ is
\begin{eqnarray}
K(m,\Delta,\mu) & = & 
\left(m^2-{2\over 3}\Delta^2\right)\log\left({m^2\over\mu^2}\right)
\ +\ 
{2\over 3}\Delta \sqrt{\Delta^2-m^2}
\log\left({\Delta-\sqrt{\Delta^2-m^2+ i \epsilon}\over
\Delta+\sqrt{\Delta^2-m^2+ i \epsilon}}\right)
\nonumber\\
& & \ +\ {2\over 3} {m^2\over\Delta} \left(\ \pi m - 
\sqrt{\Delta^2-m^2}
\log\left({\Delta-\sqrt{\Delta^2-m^2+ i \epsilon}\over
\Delta+\sqrt{\Delta^2-m^2+ i \epsilon}}\right)
\right)
\ \ \ ,
\label{eq:Kdecfun}
\end{eqnarray}
where $\mu$ is the renormalization scale.
The last contribution in eq.~(\ref{eq:qcdaxials}) is somewhat interesting as it depends 
upon the extension from QCD to PQQCD and also upon the choice of lattice
fermions and does not vanish in the isospin-symmetric QCD limit.

%%%%%%%%%%%%%%%  Twist  %%%%%%%%%%%%%%%%%%%%
\section{Matrix Elements of the Isovector Twist-2 Operators}

The forward matrix elements   of twist-2 operators play  an  important
role in hadronic structure as they are directly related to the moments
of the    parton  distribution functions.  In  QCD   the long-distance
contributions    to these  matrix    elements    have  been   computed
order-by-order in the chiral expansion using $\chi$PT~\cite{wally,AS,CJ,CJb}
and have  been applied to results  from  both quenched and unquenched
lattice  data~\cite{wally}, with  interesting results.  Further, the
analogous contributions in Q$\chi$PT and PQ$\chi$PT have been computed
in Refs.~\cite{CSn,BSn,Chen:2001gr}.

In QCD, in the continuum limit the twist-2 operators are 
\begin{eqnarray}
{\cal O}^{ (n), b}_{\mu_1\mu_2\ ... \mu_n}
& = & 
\overline{q}\ \tau^b\ \gamma_{ \mu_1  } 
\left(i \stackrel{\leftrightarrow}{D}_{\mu_2}\right)\ 
... 
\left(i \stackrel{\leftrightarrow}{D}_{ \mu_n }\right)\ q
\ -\ {\rm traces}
\ \ \ ,
\label{eq:twistop}
\end{eqnarray}
where it is understood that the operator is symmetrized on its Lorentz indices.
For the forward matrix elements in the proton and neutron, as are
relevant for deep inelastic scattering, we only need to consider
$b=3$.  The extension to PQQCD is straightforward,
\begin{eqnarray}
^{(PQ)}{\cal O}^{ (n), b}_{\mu_1\mu_2\ ... \mu_n}
& =& 
\overline{Q}\ \overline{\tau}^b\ \gamma_{ \mu_1  } 
\left(i \stackrel{\leftrightarrow}{D}_{\mu_2}\right)\ 
... 
\left(i \stackrel{\leftrightarrow}{D}_{ \mu_n }\right)\ Q
\ -\ {\rm traces}
\ \ \ ,
\label{eq:twistopPQ}
\end{eqnarray}
where $\overline{\tau}^b$ has the same form as that in 
eq.~(\ref{eq:axcharge}), and we will use the same charges, but it should be
remembered that they are unrelated to those of the axial-current operators.

The higher-dimensional operators that enter at finite lattice spacing
are, in general, quite complicated.  This is due to that fact that the
operators must be classified not only by charge conjugation, parity
and so forth, but also by the representation theory of the hypercubic group,
$H(4)$.  Such a classification has been performed up to
$n=4$~\cite{Gockeler:1996mu}, but as $n$ increases the number and
complexity of the operator basis increases significantly.

The $n=1$ operator, ${\cal O}^{ (1), 3}_{\mu}$, is the isovector vector-current operator
with ${\cal O}(a)$ corrections~\cite{Capitani:2000xi}
\begin{eqnarray}
\overline{q}\ \tau^3\  \gamma^\mu\  m_q\  q
\ \ \ , \ \ \ 
\partial_\nu\ \left(\ \overline{q}\ \tau^3\ \sigma^{\mu\nu}\ q\ \right)
\ \ \ .
\end{eqnarray}
The forward matrix element of the second operator vanishes for obvious
reasons and the first operator is ${\cal O}(p^4)$ in the expansion,
and thus there are no operator corrections 
to ${\cal O}^{ (1), 3}_{\mu}$
to the order we are
working.  Therefore, there are no modifications to the matrix element
of ${\cal O}^{ (1), 3}_{\mu}$.

By contrast, the forward matrix element of the $n=2$ operator, 
\begin{eqnarray}
{\cal O}^{ (2), 3}_{\mu\nu}
& = & 
\overline{q}\ \tau^3\ 
\gamma_\mu  \left(i \stackrel{\leftrightarrow}{D}_{\nu}\right)\ q
\ \ \ ,
\label{eq:twistn2}
\end{eqnarray}
does receive corrections at ${\cal O}(p^2)$~\cite{Capitani:2000xi}.
The correction to ${\cal O}^{ (2), 3}_{\mu\nu}$ at ${\cal O}(p^2)$
in QCD is
\begin{eqnarray}
\delta {\cal O}^{ (2), 3}_{\mu\nu}
& = & 
a c_1^{(2)}\ \overline{q}\ \tau^3\ 
\sigma_{\mu\lambda}  \left[\ i \stackrel{\leftrightarrow}{D}_{\nu} \ ,\  
i \stackrel{\leftrightarrow}{D}_{\lambda}\ \right]\ q
\ +\ 
a c_2^{(2)}\ \overline{q}\ \tau^3\ 
\{\ i \stackrel{\leftrightarrow}{D}_{\mu} \ ,\  
i \stackrel{\leftrightarrow}{D}_{\nu}\ \}\ q
\ \ \ .
\label{eq:twistn2a}
\end{eqnarray}
When extended to PQQCD, this correction becomes
\begin{eqnarray}
^{(PQ)}\delta{\cal O}^{ (2), 3}_{\mu\nu}
& = & 
\overline{Q}\ \overline{\tau}^3_{A,1}\ 
\sigma_{\mu\lambda}  \left[\ i \stackrel{\leftrightarrow}{D}_{\nu} \ ,\  
i \stackrel{\leftrightarrow}{D}_{\lambda}\ \right]\ Q
\ +\ 
\overline{Q}\ \overline{\tau}^3_{A,2}\ 
\{\ i \stackrel{\leftrightarrow}{D}_{\mu} \ ,\  
i \stackrel{\leftrightarrow}{D}_{\nu}\ \}\ Q
\ \ \ ,
\label{eq:twistn2aPQ}
\end{eqnarray}
where the charge matrices are
\begin{eqnarray}
\overline{\tau}^3_{A,1} & = & 
a \ {\rm diag}\left( c_1^{(2)(V)} , -c_1^{(2)(V)} , y_j\  c_1^{(2)(S)} , 
y_l \ c_1^{(2)(S)} , y_j\  c_1^{(2)(V)} , y_l \ c_1^{(2)(V)}\ \right) \ \ ,
\nonumber\\
\overline{\tau}^3_{A,2} & = & 
a \ {\rm diag}\left( c_2^{(2)(V)} , -c_2^{(2)(V)} , y_j\  c_2^{(2)(S)} , 
y_l \ c_2^{(2)(S)} , y_j\  c_2^{(2)(V)} , y_l \ c_2^{(2)(V)}\ \right)
\ \ \ ,
\end{eqnarray}
for the charge matrix given in eq.~(\ref{eq:axcharge}).
The coefficients $ c_{1,2}^{(2)(V)}$ and $ c_{1,2}^{(2)(S)}$
are the valence- and sea-quark coefficients, respectively.
In order to construct the nucleon matrix elements we define the fields
\begin{eqnarray}
\overline{\tau}^{(2),3}_{A,1,\xi+} & = &
\ {1\over 2}\left(\ 
\xi\overline{\tau}^3_{A,1}\xi + \xi^\dagger\overline{\tau}^3_{A,1}\xi^\dagger\
\right)
\ \ ,\ \ 
\overline{\tau}^{(2),3}_{A,2,\xi+} \ =\ 
\ {1\over 2}\left(\ 
\xi\overline{\tau}^3_{A,2}\xi + \xi^\dagger\overline{\tau}^3_{A,2}\xi^\dagger\
\right)
\ \ \ .
\end{eqnarray}

In QCD, the matrix elements of the continuum operator
${\cal O}^{ (2), 3}_{\mu\nu}$ are reproduced 
at leading order by 
\begin{eqnarray}
{\cal O}^{(2),3}_{\mu\nu} & & \rightarrow 
\rho^{(2)}\ v_{\mu} v_{\nu}\ 
\overline{N} \tau^3_{\xi +} N
\ + \   
\gamma^{(2)} 
\ v_{\mu} v_{\nu}\ 
\overline{T}^\alpha\  \tau^3_{\xi +}\ T_\alpha
\ +\  
\sigma^{(2)} {1\over 2}\left[\ 
\overline{T}_{\mu}\ \tau^3_{\xi +}\ T_{\nu}
\ +\ \overline{T}_{\nu}\ \tau^3_{\xi +}\ T_{\mu}
\ \right]
\nonumber\\ & & 
\ -\ {\rm traces}
\ \ \ ,
\label{eq:treeQCD}
\end{eqnarray}
and in PQQCD this becomes
\begin{eqnarray}
^{PQ}{\cal O}^{(2),3}_{\mu\nu}
& \rightarrow  &
\alpha^{(2)}_0\ v_{\mu} v_{\nu}\ 
\left(\ \overline{\cal B}\  {\cal B}\  \overline{\tau}^3_{\xi +}\ \right)
\ +\ 
\beta^{(2)}_0\ v_{\mu} v_{\nu}\ 
\left(\ \overline{\cal B}\  \overline{\tau}^3_{\xi +}\  {\cal B}\ \right)
\ +\ 
\gamma^{(2)}_0 
\ v_{\mu} v_{\nu}\ 
\left(\ \overline{\cal T}^\alpha\  \overline{\tau}^3_{\xi +} \ {\cal T}_\alpha \right)
\nonumber\\
& + & 
\sigma^{(2)}_0 {1\over 2}\ \left[\ 
\left(\ \overline{\cal T}_{\mu}\  \overline{\tau}^3_{\xi +}\ 
{\cal T}_{\nu} \right)
\ +\ 
\left(\ \overline{\cal T}_{\nu}\  \overline{\tau}^3_{\xi +}\ 
{\cal T}_{\mu} \right)
\ \right]
\ -\ {\rm traces}
\ \ \ .
\label{eq:tree}
\end{eqnarray}

The ${\cal O}(a)$ corrections to the operator, give rise to 
${\cal O}(p^2)$ corrections to the nucleon matrix elements of the form
\begin{eqnarray}
^{PQ}\delta{\cal O}^{{\cal O},(2),3}_{\mu\nu}
& =  &
\alpha^{(2)}_{11}\ v_{\mu} v_{\nu}\ 
\left(\ \overline{\cal B}\  {\cal B}\  \overline{\tau}^{(2),3}_{A,1,\xi+}\ \right)
\ +\ 
\beta^{(2)}_{11}\ v_{\mu} v_{\nu}\ 
\left(\ \overline{\cal B}\  \overline{\tau}^{(2),3}_{A,1,\xi+}\  {\cal B}\
\right)
\nonumber\\
& + & 
\alpha^{(2)}_{12}\ v_{\mu} v_{\nu}\ 
\left(\ \overline{\cal B}\  {\cal B}\  \overline{\tau}^{(2),3}_{A,2,\xi+}\ \right)
\ +\ 
\beta^{(2)}_{12}\ v_{\mu} v_{\nu}\ 
\left(\ \overline{\cal B}\  \overline{\tau}^{(2),3}_{A,2,\xi+}\  {\cal B}\
\right)
\ -\ {\rm traces}
\ \ \ ,
\label{eq:treeOPA}
\end{eqnarray}
while the contributions from a single insertion of the light-quark
mass matrix and from the Sheikholeslami-Wohlert term
at ${\cal O}(p^2)$ are
\begin{eqnarray}
^{PQ}\delta{\cal O}^{\Gamma,(2),3}_{\mu\nu}
& =  &
2 \left[\ 
b_{1, \Gamma }^{(2)}\  \cbb^{kji}\ \{\  \overline{\tau}^a_{\xi +}\ ,\ 
\Gamma_+\ \}^n_i\ \cb_{njk}
\right.\nonumber\\ & & \left.
+\ 
b_{2, \Gamma }^{(2)}\ (-)^{(\eta_i+\eta_j)(\eta_k+\eta_n)}\ 
\cbb^{kji}\ \{\  \overline{\tau}^a_{\xi +}\ ,\ \Gamma_+\ \}^n_k\ \cb_{ijn}
\right.\nonumber\\ & & \left.
+\ 
b_{3, \Gamma }^{(2)}\  (-)^{\eta_l (\eta_j+\eta_n)}\
\cbb^{kji}\  \left(\overline{\tau}^a_{\xi +}\right)^l_i\ 
\left( \Gamma_+\right)^n_j \ \cb_{lnk}
\right.\nonumber\\ & & \left.
+\ 
b_{4, \Gamma }^{(2)} \  (-)^{\eta_l \eta_j + 1}\ 
\cbb^{kji}\ \left(  
\left(\overline{\tau}^a_{\xi +}\right)^l_i\ \left( \Gamma_+\right)^n_j
\ +\ \left(\Gamma_+\right)^l_i 
\left(\overline{\tau}^a_{\xi +}\right)^n_j \right)\ \cb_{nlk}
\right.\nonumber\\ & & \left.
+\ b_{5, \Gamma }^{(2)}\  (-)^{\eta_i(\eta_l+\eta_j)}\ 
\cbb^{kji} \left(\overline{\tau}^a_{\xi +}\right)^l_j 
\left( \Gamma_+\right)^n_i\ \cb_{nlk}
\ +\ b_{6, \Gamma }^{(2)}\  \cbb^{kji}  \left(\overline{\tau}^a_{\xi +}\right)^l_i 
\ \cb_{ljk}
\ {\rm str}\left( \Gamma_+ \right) 
\right.\nonumber\\ & & \left.
\ +\ b_{7, \Gamma }^{(2)}\  \ (-)^{(\eta_i+\eta_j)(\eta_k+\eta_n)}\ 
\cbb^{kji}  \left(\overline{\tau}^a_{\xi +}\right)^n_k \ \cb_{ijn}
\ {\rm str}\left( \Gamma_+ \right) 
\right.\nonumber\\ & & \left.
\ +\ b_{8, \Gamma }^{(2)}\ \cbb^{kji}\ \cb_{ijk} 
\ {\rm str}\left(\overline{\tau}^a_{\xi +}\   \Gamma_+ \right) 
\ \ \right]
\ v_{\mu} v_{\nu} \ -\ {\rm traces}
\ ,
\label{eq:DIScts}
\end{eqnarray}
where $\Gamma={\cal M}$ and ${\cal A}$.

The forward matrix elements between nucleon states can be written 
as~\cite{BSn} 
\begin{eqnarray}
\langle ^{PQ}{\cal O}^{(2),3}_{\mu\nu} \rangle_i
& = &
\left[\ 
\rho_i^{(2)}
\ +\ {1\over 16\pi^2 f^2}
\left(\ 
\eta_i^{(2), 0}\ -\ \rho_i^{(2)} w_i\ +\ y_{j}\ \eta^{(2), j}_i
\ +\ y_l\ \eta^{(2), l}_i
\ \right)
\right.\nonumber\\ & & \left.\qquad
\ +\ c_i^{(2), 0} +\ y_{j}\ c^{(2), j}_i
\ +\ y_l\ c^{(2), l}_i
\ +\ d_i^{(2), 0} +\ y_{j}\ d^{(2), j}_i
\ +\ y_l\ d^{(2), l}_i
\ \right] 
\nonumber\\ & &
\overline{U}_i\  v_{\mu} v_{\nu}\ U_i
\ -\ {\rm traces}
\ ,
\label{eq:ttmat}
\end{eqnarray}
where expressions for
$\rho_i^{(2)}$, $\eta_i^{(2), 0}$, $ w_i$, $\eta^{(2), k}_i$,
$c_i^{(2), 0}$ and the $c^{(2), k}_i$ can be found in 
Ref.~\cite{BSn} with the understanding that the meson masses are evaluated at
finite lattice spacing according to eq.~(\ref{eq:mesmasses}).
The constants $d_i^{(2), 0}$ and $d^{(2), k}_i$ arise from 
the operators
$^{PQ}\delta{\cal O}^{{\cal O},(2),3}_{\mu\nu}$,
$^{PQ}\delta{\cal O}^{{\cal M},(2),3}_{\mu\nu}$ and
$^{PQ}\delta{\cal O}^{{\cal A},(2),3}_{\mu\nu}$,
and are found to be 
\begin{eqnarray}
d_p^{(2), 0} & = & 
{a c_1^{(2)(V)}\over 3} \left( 2 \alpha_{11}^{(2)} - \beta_{11}^{(2)}\ \right)
\ +\ 
{a c_2^{(2)(V)}\over 3} \left( 2 \alpha_{12}^{(2)} - \beta_{12}^{(2)}\ \right)
\nonumber\\
&& \ +\ 
{1\over 3} a c_{sw}^{(V)}\left(
\ -2 b_{1,A}^{(2)}+4 b_{2,A}^{(2)} - b_{3,A}^{(2)}+ b_{4,A}^{(2)}
+2 b_{5,A}^{(2)}
\right)
\ +\ 
{2\over 3} a c_{sw}^{(S)}\left( - b_{6,A}^{(2)} + 2 b_{7,A}^{(2)} \ \right) \ \ ,
\nonumber\\
d^{(2), j}_p & = & d^{(2), l}_p\ =\ 
\left(\ a c_{sw}^{(S)} - a c_{sw}^{(V)}\ \right)  b_{8,A}^{(2)}
\ \ \ ,
\end{eqnarray}
and those for the neutron are related by
$d_n^{(2), 0}  =  -d_p^{(2), 0}$ and 
$d^{(2), j}_p  =  d^{(2), j}_n  =  d^{(2), l}_n$.

In the isospin-symmetric QCD limit at finite lattice spacing these expressions reduce to
\begin{eqnarray}
\langle ^{PQ}{\cal O}^{(2),3}_{\mu\nu} \rangle_p
& = & 
\left[ \rho_p^{(2)}\left( 1 - {(3 g_A^2+1)\over 8\pi^2 f^2}L_\pi
\right)
\ -\ {g_{\Delta N}^2\over 4\pi^2 f^2} J_\pi 
\left[ \rho_p^{(2)} + {5\over 9}\gamma^{(2)} - {5\over 27}\sigma^{(2)}\right]
\right.\nonumber\\ &&\left.
\ +\ {\overline{m}\over 3} \left( -2 b_{1,M}^{(2)} + 4  b_{2,M}^{(2)}
-  b_{3,M}^{(2)} +  b_{4,M}^{(2)} + 2  b_{5,M}^{(2)} 
- 2  b_{6,M}^{(2)} + 4  b_{7,M}^{(2)}\ \right)
\right.\nonumber\\ &&\left.
\ +\ 
{a c_1^{(2)(V)}\over 3} \left( 2 \alpha_{11}^{(2)} - \beta_{11}^{(2)}\ \right)
\ +\ 
{a c_2^{(2)(V)}\over 3} \left( 2 \alpha_{12}^{(2)} - \beta_{12}^{(2)}\ \right)
\right.\nonumber\\ &&\left.
\ +\ 
{1\over 3} a c_{sw}^{(V)}\left(
\ -2 b_{1,A}^{(2)}+4 b_{2,A}^{(2)} - b_{3,A}^{(2)}+ b_{4,A}^{(2)}
+2 b_{5,A}^{(2)}
\right)
\ +\ 
{2\over 3} a c_{sw}^{(S)}\left( - b_{6,A}^{(2)} + 2 b_{7,A}^{(2)} \ \right)
\right.\nonumber\\ &&\left.
\ +\  (y_j+y_l) \left(\ a c_{sw}^{(S)} - a c_{sw}^{(V)}\ \right)  b_{8,A}^{(2)}
\right]\overline{U}_p\  v_{\mu}v_{\nu} U_p\ -\ {\rm traces}
\ \ \ .
\label{eq:ntwistQCD}
\end{eqnarray}
Like the matrix elements of the axial current, the choice of charges in the
sea quark and ghost sectors remain at finite lattice spacing through the last
term in eq.~(\ref{eq:ntwistQCD}) with coefficient $b_{8,A}^{(2)}$.

%%%%%%%%%%%%%%%%%%  Conclusions  %%%%%%%%%%%%%%%%%%%
\section{Conclusions}

As lattice QCD moves closer to its ultimate goal of computing 
strong-interaction observables directly from first principles, effective
field theory calculations must be developed in parallel in order to
facilitate comparison with data and to make rigorous predictions.
Significant attention has been paid to the chiral extrapolation of
existing quenched and unquenched data with the goal of 
making a connection between
lattice calculations performed at unphysically large quark masses and
nature.  With this program maturing pleasantly, the time is ripe to address
other extrapolations that need to be performed in order to
make a rigorous connection with data: the continuum extrapolation,
$a\rightarrow 0$, and the infinite volume extrapolation.

In this work we have computed the leading effects of a finite lattice
spacing, at ${\cal O}(a)$, on some nucleon properties.  One source of
this dependence is the leading ${\cal O}(a)$ corrections to the
strong-interaction Lagrange density, i.e. the Sheikholeslami-Wohlert
term.  However, when considering matrix elements of operators, there
are additional contributions from the ${\cal O}(a)$ corrections to the
operators themselves.  If the lattice calculations are performed with
lattice fermions that respect chiral symmetry then all of the finite
lattice spacing contributions we have computed in this work will
vanish.  This will also be the case for ${\cal O}(a)$-improved lattice
simulations. The continuum extrapolation of unimproved simulations of
nucleon properties with Wilson quarks or mixed quarks,
e.g. Ginsparg-Wilson valence quarks and Wilson sea quarks, can be
performed with the expressions we have determined in this work.

\bigskip\bigskip

\acknowledgements

We thank Steve Sharpe for many useful discussions.  
This work is supported in part by the 
U.S. Dept. of Energy under Grant No.~DE-FG03-97ER4014
(M.J.S.) and Grant No.~DE-FG03-00-ER-41132 (S.R.B.).

\end{document}